\documentclass[%
 reprint,
 amsmath,amssymb,
 aps,
]{revtex4-1}

\usepackage{graphicx}
\usepackage{dcolumn}
\usepackage{bm}

\begin{document}

\title{Dark Energy Gravitational Wave Observations and Ice Age Periodicity}

\author{Anupam Singh}
 
 \affiliation{Physics Department, L.N. Mittal I.I.T, Jaipur, Rajasthan, India.}

\date{\today}

\begin{abstract}

Dark Energy is the dominant component of the energy density of the Universe.
 However, it is also very elusive since its interaction with the rest of the Universe is primarily gravitational.
Since Dark Energy is a low energy phenomenon from the perspective of particle physics and field theory, a fundamental approach based on fields in curved space is sufficient to understand the current dynamics of Dark Energy. The key issue is to understand the gravitational dynamics of Dark Energy and its observational consequences.
However, finding the observational consequences of Dark Energy dynamics has been a very challenging task. For something which is the dominant component of the energy density of the Universe, Dark Energy appears to be very distant and reclusive. 
Here we show that the Dark Energy dynamics results in the production of gravitational waves which produce the ellipticity variation in earth’s orbit that results in the periodicity of the Ice Ages observed and documented  by geologists and climatologists.
Previously, no observational signature of gravitational waves produced by Dark Energy dynamics has been reported. Further, no interpretation of the ellipticity variation of the earth’s orbit due to gravitational waves or the linking of such gravitational waves to the Ice Age periodicity has been reported previously. We hope that the current work will lead to some fresh insights and some more interesting work.

\end{abstract}


\maketitle

\section{INTRODUCTION}


An increasing accuracy of astronomical observations has led us to an
increased precision in the determination of cosmological parameters.
This, in turn, led us to critically re-examine our cosmological models.
In particular,  the accurate determination of the Hubble constant and
the independent determination of the age of the universe forced us to
critically re-examine the simplest cosmological model
- a flat universe with a zero cosmological constant\cite{Pierce,Freedman}.
These observations forced us\cite{Singh} to consider the idea of
a small non-vanishing vacuum energy due to fields as playing an important role in the Universe.
Subsequently, there has been a large body of work both on the observational and theoretical side
that has firmed up our belief in what we now call dark energy. For a thorough discussion on Dark Energy and related issues, please see  the excellent review \cite{MarkAlessandra} .

Thus, it is important to understand the dynamics of dark energy -  in particular the gravitational dynamics of dark energy and its observational consequences.

Before we start our detailed look at the gravitational dynamics of dark energy field configurations we would like to
quickly introduce field theory models of Dark Energy so that all readers can readily relate to the discussion that follows.

Discussions of field theory models for dark energy and connections to particle physics were discussed in detail earlier \cite{Singh}.
It was noted in that article that these fields must have very light mass scales in order to be cosmologically relevant today.
Discussions of realistic particle physics models with particles of light masses capable of generating interesting cosmological consequences
have been carried out by several authors\cite{GHHK}. It has been
pointed out that the most natural class of models in which to realize these ideas are models of neutrino masses with 
Pseudo Nambu Goldstone Bosons (PNGB's). The reason for this is that the mass scales associated with such
models can be related to the neutrino masses, while any tuning that needs to be
done is protected from radiative corrections by the symmetry that gave rise
to the Nambu-Goldstone modes\cite{'thooft}.

Holman and Singh\cite{HolSing} studied the finite temperature behavior of the
see-saw model of neutrino masses and found phase transitions in this model
which result in the formation of topological defects.  In fact, the critical
temperature in this model is naturally linked to the neutrino masses.

Today we know that Dark Energy is an important component of the Universe. Thus it is natural to expect that Dark Energy will play an important role in different aspects of the Universe. In previous papers, it has been shown that Dark Energy can play an important role in structure formation and the physics of the galaxy \cite{DarkEnergyCollapseAndBHs}. In particular, the dynamical length scale associated with the Dark Energy is comparable to the galactic length scale. Also, Dark Energy collapse can result in the formation of supermassive black holes with masses comparable to the mass of the Black Hole at the center of our galaxy. 

In our quest for interactions of the Dark Energy with the rest of the Universe, our first attempts perhaps justifiably focus on the gravitational interactions of the Dark Energy, since it is the dominant interaction of Dark Energy with the rest of the Universe. One new signal that one can perhaps hope to measure is the gravitational waves emitted by Dark Energy. Thus, a study was made of the gravitational waves emitted by Dark Energy\cite{DarkEnergyGravitationalWaves}. One of the key results obtained in that paper is that the time period of the gravitational waves emitted is $ \sim 10^5$ years .

At first, one may be filled with dismay that with the time period being $ \sim 10^5 $  years and the human lifetime being $ \sim 10^2  $ years, the chances of detecting such gravitational waves are slim to none. However, nature has been kind to us.

To find a pathway to the detection of these gravitational waves, it is perhaps worth recalling that a gravitational wave periodically turns circles into ellipses and vice versa as it passes through a region of space. For an easily accessible discussion, please see Gravitational Waves, Sources, and Detectors\cite{GWsourcesDetection}. Thus, we expect a periodic oscillation in the ellipticity of orbits as a signature of gravitational waves.

Nature has in fact been kind to us in that it has not only detected these gravitational waves but maintained a record of it for us to uncover and interpret.

For this, we turn to a description of Ice Age periodicity, the underlying geological data and its interpretation in the terms of the periodic oscillations in the ellipticity of earth's orbit. 

It turns out that geologists and climatologists have known for about a century that the periodicity in the ice ages can be linked to what they call insolation which essentially quantifies the heat received from the sun. This insolation has been tied to the variations in the Earth's orbit. It further turns out that the dominant driver of this phenomena is the variation in ellipticity on the timescale of $ \sim 10^5 $  years.
For more details on this please see  Variations in the Earth's Orbit: Pacemaker of the Ice Ages\cite{ScienceIceAgesPacemaker} and Climate and atmospheric history of the past 420,000 years from the
Vostok ice core, Antarctica\cite{NatureClimaticHistoryAndEccenttricity} and refernces therein. These articles bring out the role played by periodic variations in the eccentricity of the Earth's orbit on the time scale of $ \sim 10^5 $  years in driving the periodicity of the Ice Ages.

In the following sections of the paper, we show how how one can do the time evolution of the Dark Energy fields in the presence of gravity and extract the gravitational wave signals. Finally, we show that the gravitational wave signal produced by the Dark Energy fields can produce the ellipticity variations in the earths orbit resulting in the observed periodicity of the Ice Ages.

\section{  Gravitational Dynamics  of the Dark Energy Fields}

We now quickly re-visit the study of the gravitational dynamics of Dark Energy field configurations.
The dynamics of fields in cosmological space-times has been extensively discussed
elsewhere (see e.g. Kolb and Turner\cite{rockymike} ). Likewise, gravitational collapse
in the context of general relativity has also been extensively discussed elsewhere 
(see e.g. Weinberg\cite{weinberggr}). These ideas can be pulled together to write 
down the evolution equations describing the coupled dynamics of the field configurations and
space-time interacting with each other.

We use the 3+1 BSSN formalism to numerically study the time evolution of scalar fields in the presence of gravity.
The formalism for doing this has been previously described by Balakrishna et.al.\cite{BalakrishnaEtAl}

\label{sec:evoleqns}

The action describing a self-gravitating complex scalar field in a 
curved spacetime is:

\begin{eqnarray}
I = \int d^4 x \sqrt{-g} \left( \frac{1}{16 \pi}R \, 
    -\frac{1}{2} [ g^{\mu \nu} 
    \partial_{\mu} \Phi^* \, \partial_{\nu} \Phi  
    + V(|\Phi|^2) ]  \right) 
\label{action}
\end{eqnarray}

\noindent where $R$ is the Ricci scalar, $g_{\mu\nu}$ is the metric 
of the spacetime, $g$ is the determinant of the metric, $\Phi$ is the 
scalar field, $V$ its potential. Varying this action 
leads to equations of motion for the entire system.
Variation with respect to 
the scalar field leads to the Klein-Gordon equation for the scalar 
field

\begin{equation}
\Phi^{; \mu} {}_{;\mu} - \frac{dV}{d|\Phi|^2}\Phi = 0 . 
\label{klein-gordon}
\end{equation} 

\noindent When the variation of Eq.\ (1) is made with respect to the metric 
$g^{\mu\nu}$, we get the Einstein's equations $G_{\mu\nu}= 8\pi T_{\mu\nu}$ . The resulting stress energy tensor is:

\begin{equation}
T_{\mu \nu} = \frac{1}{2}[\partial_{\mu} \Phi^{*} \partial_{\nu}\Phi +
\partial_{\mu} \Phi \partial_{\nu}\Phi^{*}] -\frac{1}{2}g_{\mu \nu}
[\Phi^{*,\eta} \Phi_{,\eta} + V(|\Phi|^2))].
\label{stress-energy}
\end{equation}

To get numerical solutions it is convenient to use the 3+1 decomposition of Einstein's equations, 
for which the line element can be written as

\begin{equation}
   d s^{2}  = - \alpha^2  dt^2 + \gamma_{ij}  (dx^i + \beta^{i} dt) (dx^j 
              + \beta^{j} dt)  
\label{line-element}
\end{equation}

\noindent
where $\gamma_{ij}$ is the 3-dimensional metric. The latin 
indices label the three spatial coordinates. The functions $ \alpha $ 
and $ \beta^{i}$ in Eq. (\ref{line-element}) are gauge 
parameters, known as the lapse function and the shift vector respectively. 
The determinant of the 3-metric is $\gamma$ . The Greek indices run from 0 to 3 and the 
Latin indices run from 1 to 3. 

For the purpose of doing numerical evolution, the Klein-Gordon equation can be written as a first-order system.
This  is done by first splitting the scalar field into the real and imaginary parts as: 
$\Phi = \phi_1 + i\phi_2$, and then defining the following variables in terms 
of combinations of their derivatives:  $\Pi = \pi_1 + i \pi_2$ and $ 
\psi_a = \psi_{1a} + i \psi_{2a}$. Here $ \pi_1= (\sqrt{\gamma}/\alpha) 
(\partial_t \phi_1 - \beta^c \partial_c \phi_1) $ and  
$\psi_{1a}=\partial_a \phi_1$ and similarly we can replace the subscript $1$ with $2$ to get the remaining quantities of interest. With this 
notation the evolution equations become 

\begin{eqnarray}
 \partial_t \phi_1 &=&  \frac{\alpha}{\gamma^{\frac{1}{2}}} \pi_1 + 
\beta^j \psi_{1j} \\\nonumber
 \partial_t \psi_{1a} &=& \partial_a( \frac{\alpha}{\gamma^{\frac{1}{2}}} 
\pi_1 + \beta^j \psi_{1j}) \\\nonumber
 \partial_t \pi_1 &=& \partial_j (\alpha \sqrt{\gamma} \phi_1^j) 
  - \frac{1}{2} \alpha  \sqrt{\gamma} \frac{\partial V}{\partial 
  \vert \Phi \vert^2} \phi_1 \label{FirstOrderKleinGordon}
\end{eqnarray} 

\noindent
and again, we can replace the subscript $1$ with $2$ to get the remaining quantities of interest.
On the other hand, the geometry of the spacetime is evolved 
using the BSSN formulation of the 3+1 decomposition. According to this 
formulation, the variables to be evolved are 
$\Psi = \ln(\gamma_{ij} \gamma^{ij})/12$, 
$\tilde{\gamma}_{ij} = e^{-4\Psi}\gamma_{ij}$, $K = \gamma^{ij}K_{ij}$, 
$\tilde{A}_{ij}=e^{-4\Psi}(K_{ij}-\gamma_{ij} K/3)$ and 
the contracted Christoffel symbols 
$\tilde{\Gamma}^{i}=\tilde{\gamma}^{jk}\Gamma^{i}_{jk}$,
instead of the ADM variables $\gamma_{ij}$ and $K_{ij}$. The 
equations for the BSSN variables are described in Refs. \cite{bssn, 
stu}:

\begin{eqnarray}
\partial_t \Psi &=& - \frac{1}{6} \alpha K \label{BSSN-MoL/eq:evolphi}\\
\partial_t \tilde{\gamma}_{ij} &=& - 2 \alpha \tilde{A}_{ij}
\label{BSSN-MoL/eq:evolg} \\
\partial_t K &=& - \gamma^{ij} D_i D_j \alpha  + \alpha \left[
        \tilde{A}_{ij} \tilde{A}^{ij} + \frac{1}{3} K^2 + \frac{1}{2}
        \left( -T^{t}{}_{t} + T \right) \right]
\label{BSSN-MoL/eq:evolK}\\
\partial_t \tilde{A}_{ij} &=& e^{-4 \Psi} \left[
 - D_i D_j \alpha + \alpha \left( R_{ij} - T_{ij} \right) \right]^{TF}
\noindent\\
&& + \alpha \left( K \tilde{A}_{ij} - 2 \tilde{A}_{il}\tilde{A}_j^l\right) \label{BSSN-MoL/eq:evolA}\\\frac{\partial}{\partial t}\tilde \Gamma^i
&=& - 2 \tilde A^{ij} \alpha_{,j} + 2 \alpha \Big(\tilde \Gamma^i_{jk} \tilde A^{kj}                              \nonumber \\
&& - \frac{2}{3} \tilde \gamma^{ij} K_{,j}- \tilde \gamma^{ij} T_{j t} + 6 \tilde A^{ij} \phi_{,j} \Big)
                                                                \nonumber \\
&& - \frac{\partial}{\partial x^j} \Big(\beta^l \tilde \gamma^{ij}_{~~,l}- 2 \tilde \gamma^{m(j} \beta^{i)}_{~,m}+ \frac{2}{3} \tilde \gamma^{ij} \beta^l_{~,l} \Big) .
\label{BSSN-MoL/eq:evolGamma2}
\end{eqnarray}

\noindent where $D_i$ is the covariant derivative in the spatial
hypersurface, $T$ is the trace of the stress-energy 
tensor~(\ref{stress-energy}) and the label $TF$ denotes the trace-free part 
of the quantity in brackets.

The above equations are true for any general potential $V$. 
One can of course write down the corresponding equations for PNGB fields.
All we need to do is  specify the appropriate potential.
In our case, the field is a real scalar field.
The simplest potential one can write down for the physically motivated PNGB fields \cite{GHHK} can be written in the form:

\begin{equation}
V = m^4 \left[ K - \cos(\frac{\Phi}{f} ) \right]
\end{equation}

As discussed in \cite{GHHK} m is of order the neutrino mass and K is of order 1. For the sake of definiteness, in what follows we will choose $K = 1 $. We will consider such a potential for studying the dynamics in the next section.

Before we turn to a discussion of the gravitational wave signal and the numerical results obtained, we identify the fundamental timescale of the dynamics as per the model described here. Of course, both the dynamics of fields and the fundamental timescale have been extensively studied before. Please see for example \cite{BrandenbergerEtAl,LindeEtAl,nonequilibrium,dissipation} for detailed discussions on this issue. It has been noted in these analyses that even in complex non-equilibrium situations the inverse mass of the field $m_{\Phi}^{-1}$ plays an important role in determining the dynamics of fields. In particular, resonances associated with this fundamental timescale play an important role in the dynamics. We will elaborate on the role of resonances in the final section of this article. For now, we focus on arriving at the fundamental timescale associated with the dynamics. To do this, we note that the mass of the field is determined by the second derivative of the field, $m_{\Phi}^2 \simeq V^{''} (\Phi)$.
The fundamental timescale of the dynamics is then given by $t \simeq m_{\Phi}^{-1}$.

For our potential given above, taking the second derivative gives us: $m_{\Phi}^2 \simeq m^4/f^2$.
Thus, the fundamental timescale of the dynamics is given by $t \simeq f/m^2$.

To convert into physical units, we note the following. 

The scale $f$ is the high energy symmetry breaking scale in PNGB models. In the see-saw model of neutrino masses\cite{Singh} this corresponds to the heavy scale of symmetry breaking. While $f$ has a range of possible values, the typical value of $f$ in the see-saw model of neutrino masses is $f \sim 10^{13} GeV$. The typical value of $m$ is given by $m \sim 10^{-3} eV $. It should also be noted that so far we have been working in the Particle Physics and Cosmology units in which $ \hbar=c=k=1 $ . It is straightforward to convert from these units into more familiar units using standard conversion factors\cite{rockymike}. Thus, $ 1 GeV^{-1} = 1.98 \times 10^{-14} cm$ and $ 1 GeV^{-1} = 6.58 \times 10^{-25} sec$. 

Using these conversion factors we see that the fundamental time scale of the Dark Energy dynamics corresponding to $\frac{f}{m^2}$ is $\sim  10^5 $ years. As we shall see this fundamental timescale plays a key role in the production of gravitational waves due to dark energy dynamics. To complete this discussion we now turn to the discussion of gravitational waves.

\section{Gravitational Wave Signal }

We are interested in the gravitational wave signal produced by the dynamics of the dark energy field whose time evolution we have described in the previous section.
This can be extracted numerically and we have used the publicly available Einstein Toolkit for this purpose.

 Here we show how to extract the gauge-invariant, odd and even perturbations. 
Background  material on this can be found in Refs. \cite{Camarda99, Rezzolla99, Baker2000}.
 
 The gravitational waves can be
considered as a  perturbation to a fixed background and we can write
\begin{equation}
g_{\mu\nu}=g^0_{\mu\nu}+h_{\mu\nu}\,,
\end{equation}
where $g^0_{\mu\nu}$ is the fixed background metric
and $h_{\mu\nu}$ its  perturbation.
The background metric $g^0_{\mu\nu}$ is
usually assumed to be of Minkowski or Schwarzschild form, which we can write as
\begin{equation}
ds^2=-N dt^2 + A dr^2 + r^2(d\theta^2 + \sin^2\theta d\phi^2)\,.
\end{equation}

We can split the spacetime into timelike ,radial
and angular parts which in turn will help us in decomposing
the metric perturbation $h_{\mu\nu}$ into odd
and even multipoles, i.e.,~we can write
\begin{equation}
h_{\mu\nu}=\sum_{\ell m}\left[\left(h_{\mu\nu}^{\ell m}\right)^{(o)} + \left(h_{\mu\nu}^{\ell m}\right)^{(e)}\right]\ .
\end{equation}

It is also possible to expand  these components in their vector and tensor spherical harmonics.

The solutions formed by odd-even parity perturbations are given by
the Regge-Wheeler-Moncrief and the Zerilli-Moncrief master functions, respectively.
The odd-parity Regge-Wheeler-Moncrief function reads
\begin{eqnarray} \label{eq:Qodd}
Q^{\times}_{\ell m} &\equiv& \sqrt{\frac{2(\ell+1)!}{(\ell-2)!}}
	\frac{1}{r}\left(1-\frac{2M}{r}\right) \nonumber \\
	& & \left[(h_{1}^{\ell m})^{({\rm o})}+\frac{r^2}{2} \partial_r
	\left(\frac{(h_2^{\ell m})^{({\rm o})}}{r^2}\right)\right]\ ,
\end{eqnarray}
and the even-parity Zerilli-Moncrief function reads
\begin{eqnarray} \label{eq:Qeven}
{Q}^{+}_{\ell m} \equiv \sqrt{\frac{2(\ell+1)!}{(\ell-2)!}}
\frac{r q_1^{\ell m}}{\Lambda\left[r\left(\Lambda-2\right)+6M\right]} \,,
\end{eqnarray}
where $\Lambda=\ell(\ell+1)$, and where
\begin{equation}
\label{q1}
q_1^{\ell m}  \equiv r\Lambda\kappa_1^{\ell m} + \frac{4r}{A^2}\kappa_2^{\ell m} \,,
\end{equation}
with
\begin{eqnarray}
\label{kappa1}
\kappa_1^{\ell m} & \equiv & K^{\ell m}+\frac{1}{A}\left(r\partial_r G^{\ell m}-
	\frac{2}{r}(h_1^{\ell m})^{({\rm e})}\right)\ ,\\
\label{kappa2}
\kappa_2^{\ell m} & \equiv &\frac{1}{2}\left[A H_2^{\ell m}-
	\sqrt{A} \partial_r \left(r \sqrt{A} K^{\ell m}\right)\right]\, .
\label{def:q1}
\end{eqnarray}

These master functions depend entirely on the spherical part of the metric given by the coefficients $N$ and
$A$, and
the perturbation coefficients for the
individual metric perturbation components $(h_{1}^{\ell m})^{({\rm o})}$, $(h_{2}^{\ell m})^{({\rm o})}$,
$(h_{1}^{\ell m})^{({\rm e})}$, $(h_{2}^{\ell m})^{({\rm e})}$, 
$H_0^{\ell m}$, $H_1^{\ell m}$, $H_2^{\ell m}$, $K^{\ell m}$, and $G^{\ell m}$ which
can be obtained
from any numerical spacetime by projecting out the Schwarzschild or Minkowski
background \cite{Camarda:1998wf}.
For example, the coefficient $H_2^{\ell m}$ can be obtained via
\begin{equation}
H_2^{\ell m}=\frac{1}{A}\int (g_{rr}-A) Y_{\ell m}\,d\Omega\,,
\end{equation}
where $g_{rr}$ is the radial component of the numerical metric represented in the spherical-polar coordinate basis,
$Y_{\ell m}$ are spherical harmonics, and $d\Omega$ is the surface line element of the $S^2$ extraction sphere.
The coefficient $A$ represents the spherical part of the background metric
and can be obtained by projection of the numerical metric component $g_{rr}$ on $Y_{00}$ over the extraction sphere
\begin{equation} \label{eq:H2lm}
A=\frac{1}{4\pi}\int g_{rr} d\Omega\,.
\end{equation}
Similar expressions hold for the remaining perturbation coefficients.

The odd- and even-parity master functions Eq.~(\ref{eq:Qodd}) and Eq.~(\ref{eq:Qeven}) can be
straight-forwardly related to the gravitational-wave strain and are given by
\begin{eqnarray}
\label{eq:h-Q}
h_{+}-\mathrm{i}h_{\times}&=&\frac{1}{\sqrt{2}r}\sum_{\ell,m}\left(
	Q^{+}_{{\ell m}} - \mathrm{i}\int_{-\infty}^{t}
	Q^{\times}_{{\ell m}}(t')dt'\right) \nonumber \\
	& & \;_{_{-2}}Y^{{\ell m}}(\theta,\phi)
	+ {\cal O}\left(\frac{1}{r^2}\right)\, ,
\end{eqnarray}
where ${_{-2}}Y^{{\ell m}}(\theta,\phi)$ are the spin-weight $s=-2$
spherical harmonics.

\section{Gravitational Waves from Dark Energy dynamics and Ice Age Periodicty}

We have used the 3+1 BSSN formalism for doing the numerical evolution.
The equations we described earlier were solved numerically using the publicly available Einstein Toolkit\cite{EinsteinToolkit}.
In particular, we extracted the gravitational wave signals emitted as a result of the dynamics of dark energy field configurations.
While a detailed discussion of the gravitational waves produced by Dark Energy dynamics has been given by us previously\cite{DarkEnergyGravitationalWaves}, here we focus on how these gravitational waves can explain the Ice Age periodicity.

In order to do this, we look both at the frequency and the amplitude of the gravitational waves produced.
First, let us look at the frequency of the gravitational waves produced. 
We note, that the time period of the gravitational waves produced is comparable to the timescale of the Dark Energy dynamics. This can be confirmed by examining the results and plots shown also in our earlier work\cite{DarkEnergyGravitationalWaves}.
It should also be noted that $\frac{f}{m^2}$  is the fundamental timescale of the dynamics as determined by the Dark Energy evolution equations and this is also determines the time period of the gravitational waves.

As already noted earlier, the fundamental time scale of the Dark Energy dynamics corresponds to $\frac{f}{m^2}$ is $\sim  10^5 $ years and this is also the time period of the gravitational waves produced by the Dark Energy fields.

We now note that the gravitational waves with a time period of $ \sim 10^5 $ years will produce periodic variations in the eccentricity of Earth's orbit with a time period of $ \sim 10^5 $ years. This in turn affects the amount of heat captured by the earth from the sun which will show the periodicity with the time period of $ \sim 10^5 $ years resulting in the observed periodicity of the Ice Ages as described and documented in \cite{NatureClimaticHistoryAndEccenttricity,ScienceIceAgesPacemaker} and references therein.

Moreover, we see that the amplitude of the gravitational waves produced by dark energy dynamics is sufficient to produce the periodic ellipticity variations in earth's orbit required to explain the Ice Age periodicity. In order to do this, we first draw your attention to the data on the amplitude required\cite{Amplitude}.This analysis shows that the ellipticity change is $\sim  10^{-2} $  and hence the required gravitational wave amplitude $ h \sim 10^{-2}$ . It turns out that this amplitude can be easily achieved through the Dark Energy dynamics. This can be inferred from the amplitude of the gravitational waves produced by dark energy dynamics obtained by us and reported in \cite{DarkEnergyGravitationalWaves}.

While the numerical results described above demonstrate that the frequency and  amplitude of the gravitational waves produced by dark energy dynamics can produce the periodic ellipticity variations in earth's orbit required to explain the Ice Age periodicity, we wish to elucidate these ideas further. 

In previous work on the dynamics of fields it has been noted\cite{BrandenbergerEtAl,LindeEtAl,nonequilibrium,dissipation} that resonance phenomena play a very important role in the dynamics of fields.

To translate this into the current context and gain further insight into the underlying physics being discussed here we use the equation of motion of the gravitational wave field  $h_{\mu\nu}$ and note that the primary mechanism for the transfer of energy from the Dark Energy field to the gravitational wave field is the resonance phenomenon. This will enable us to arrive at both the frequency and amplitude of the gravitational wave field as described and discussed below. 

The equation of motion of the gravitational waves in vacuum is:

\begin{equation}
\Box h_{\mu\nu}= 0 
\end{equation}

which follows from the Einstein Equation:

\begin{equation}
R_{\mu\nu} - \frac{1}{2} g_{\mu\nu} R= \kappa T_{\mu\nu}
\end{equation}

with the source term zero i.e. when  $T_{\mu\nu} = 0 $. See for example \cite{LL} for an excellent discussion on this and related topics.

When the source term is not zero, the equation for the gravitational wave components $ h_{\mu\nu} $ is given by:

\begin{equation}
\Box h_{\mu\nu}= S_{\mu\nu}
\end{equation}

The non-zero source term $  S_{\mu\nu}  $ is present for example if you have non-trivial scalar field configurations present.

For a scalar field source with the scalar field having a natural frequency $ \omega $ (related to the mass m of the scalar field), the source term can be expressed as :

\begin{equation}
S_{\mu\nu} = S_{\mu\nu}^{(0)} e^{i \omega t}
\end{equation}

Putting this into the equation for the gravitational wave componenets we get:

\begin{equation}
\Box h_{\mu\nu}= S_{\mu\nu}^{(0)} e^{i \omega t}
\end{equation}

This has the form of a driven harmonic oscillator which can be seen much more explicitly by going into Fourier space.

As is well known from the solution of the driven harmonic oscillator, this implies two things:

(1) The resonance picks out some special frequencies and enhances their amplitude. This is the reason that the time period of the Gravitational Waves matches the time period of scalar field dynamics as already observed.

(2) The amplitude at resonance can grow to be large.

Looking at the equation for the gravitational wave components in Fourier space, we get:

\begin{equation}
\frac{d^2 h_{\mu\nu}}{d t^2}  + k^2 h_{\mu\nu} = S_{\mu\nu}^{(0)} e^{i \omega t}
\end{equation}

The well known solution to the driven harmonic oscillator implies that as long as the energetics permit, the gravity wave amplitude can grow to very large values given by:

\begin{equation}
h_{\mu\nu}^{(0)} = \frac{S_{\mu\nu}^{(0)}}{(k^2 - \omega^2)}
\end{equation}

showing the typical large value at resonance as $ k \to \omega $.
Thus the mode at resonance can grow to very large values.
Energetics will prevent the amplitude from going to infinite values and will limit to large but finite values.
This is consistent with our numerical results which show that large but finite values of the gravity wave amplitude can be produced by the dark energy dynamics.

To summarize the resonance production of gravitational waves, in particular from scalar field dynamics gives us 2 key features of the gravitational waves observed by us:

(1) Because of resonance the time period of gravitational waves matches the time period of scalar field dynamics.

(2) The amplitude of the gravity waves can reach large values due to the phenomenon of resonance.

We now want to push these ideas one step further by using the energetics to arrive at an estimate of the amplitude of the gravitational waves produced by dark energy dynamics. In order to do this, we first note that the system of interest for us here consists of 2 important sub-components: namely dark energy and gravitational waves. These two sub-components can exchange energy as they interact dynamically. Initially, when the gravity wave amplitudes are very small, the direction of transfer of energy will be from the dark energy field to the gravitational wave field. However, as the gravitational wave amplitude grows as a result of the resonance phenomenon, it will also start dynamically transferring some energy back to the dark energy field. Since at this point there is a significant flow of energy in both directions (from and towards the dark energy field) the long term state of the system of interest will be dynamically driven towards the point where the energy density in gravitational waves $\rho_{gw}$ is equal to the  energy density in the dark energy field $\rho_{DE}$. Thus we can use the condition (which we henceforth refer to as the equipartition condition):

\begin{equation}
 \rho_{gw} \simeq  \rho_{DE}
\end{equation}

to determine the amplitude of the gravitational waves at long times.
One can arrive at the equipartition condition from a Statistical Mechanics perspective. 
In order to do this, we have to note from the equation of motion for the gravitational wave field  $h_{\mu\nu}$ in Fourier space, that for a given momentum, the gravitational wave field is essentially a harmonic oscillator coupled to the Dark Energy field. Thus, we are essentially dealing with a system of oscillators for which one can construct a canonical ensemble and use standard statistical mechanics tenchiques to arrive at the equipartition of energy - please see for example \cite{Reif} and references therein for additional details on this and related issues.

We now translate this into an estimate for the amplitude of the gravitational waves produced.

In order to do this we consider the observational data. We need to start with the Dark Energy density. We now know that the Dark Energy density is the dominant component of the energy density of the Universe. For estimating the gravitational wave amplitude inside our galaxy we need to start with the energy density within our galaxy. This can be estimated fairly robustly from available data (see for example \cite{MilkyWaydata}). We can summarize this data here by noting that a mass $ \sim 10^{12} M_{\odot} $ is confined to a volume with linear dimensions $\sim 10 $ kpc . From this data we can estimate that on galaxy length scales (within our galaxy) the dark energy density is $ \rho_{DE} \sim 10^{-1} $ erg/cm$^3 $.

We now turn to considering the energy density in gravitational waves and using the equipartition condiion to determining the gravitational wave amplitude.

For a gravitational wave with amplitude $ h_0 $ and angular frequency $\omega$, the energy density in gravitational waves $  \rho_{gw} $ is given by ( See for example \cite{LL} for an excellent discussion on this and related topics.):

\begin{equation}
 \rho_{gw} \simeq \frac{c^2 \omega^2  h_0^2}{(32 \pi G)}
\end{equation}

We note that the angular frequency is related to the time period $T$ by $\omega = 2 \pi / T $. Both from the resonance condition as well as the observational data on the time period $T$ we get that $T \sim 10^5 years $.
Inserting this into the equation for the gravitational wave energy density and using the equipartition condition we get
$ h_0 \sim 10^{-2} $ as the amplitude for the gravitational waves produced by the dark energy dynamics.

It is worth pointing out at this juncture that the value of the gravitational wave amplitude as calculated above  is of the correct magnitude to explain the data on the periodic eccentricity variation of earth's orbit as documented by Hinnov in the Ann. Rev. Earth Planet. Sci. (2000)\cite{Amplitude}.

Thus, we see that the resonance analysis helps us not only to intuitively understand and gain insights into the basic underlying physics, but also helps us to quantitatively arrive at the correct magnitude for both frequency and amplitude of the gravitational waves produced by the dark energy dynamics which matches the observational data on the Ice Age periodicity as given for example  by Hinnov\cite{Amplitude}.

In summary, we see that the gravitational waves produced by dark energy dynamics can produce the periodic variations in the Earth's orbital eccentricity resulting in the Ice Age periodicty.

In the current work, we have described how gravitational waves with an amplitude $ h_0 \sim 10^{-2} $  and with a time period $T \sim 10^5 years $ which corresponds to a frequency of  $ \sim10^{-13} Hz$ can arise out of dark energy dynamics.
The value of $ h_0 \sim 10^{-2} $ at the low frequencies of  $ \sim10^{-13} Hz$ is compatible with other constraints on gravitational waves - please see for example\cite{GWobservations} for a thorough discussion on existing constraints. While it is reasonable to expect the background of gravitational waves as a result of the dark energy dynamics, it is fair to say that the observation of such waves (other than through the variation of earth's orbit leading to the ice age periodicity) will be very challenging. The primary reason this will be challenging is because of the low frequency $ \sim10^{-13} Hz$. It should be noted that this corresponds to a time period $T \sim 10^5 years $ which is larger than a human lifetime by a factor of $1000$.

There has been a significant amount of interesting work done on dark energy and gravitational waves - please see for example \cite{gwde} and references therein.  It should be pointed out that none of the other works on dark energy and gravitational waves has predicted gravitational waves with the correct frequency and amplitude to help explain the ice age periodicity arising from the ellipticity variations of earth's orbit. While the other approaches to dark energy introduce many new ideas worth exploring, it is perhaps fair to note that none of them so far have attempted to incorporate the full field theory implications of known particle physics species - of these the most relevant at the current energy density of the universe would be the incorporation of neutrino physics. The approach we have taken to dark energy \cite{Singh, DarkEnergyCollapseAndBHs, DarkEnergyGravitationalWaves} incorporates the see-saw model of neutrino masses and its observational implications. This in turn through the resonance effects as described in the current work leads to the production of gravitational waves with the correct frequency and amplitude to help explain the ice age periodicity arising from the ellipticity variations of earth's orbit.

Finally, we note that dark energy is connected to many fundamental aspects of physics. Further, it has observational consequences, some of which have already been established and others which may be within our reach in our lifetimes. Because of the many connections to fundamental physics and observational consequences, dark energy is one of our best bets to uncover deep secrets of nature and push the frontiers of human knowledge. We hope that the current work will help stimulate further thought and work that may help uncover even deeper secrets of nature.


\noindent\rule{13.7cm}{0.4pt}
\newline

\centerline{\bf ACKNOWLEDGEMENTS}

This work was supported in part by the DST grant SB/S2/HEP-005/2013 from the Government of India.  

\frenchspacing

\noindent\rule{13.7cm}{0.4pt}


 \newpage

\end{document}